# Tunable ferroelectric topological defects on 2D topological surfaces: strain engineering skyrmion-like polar structures in 2D materials

Bo Xu[1,#], Zhanpeng Gong[1,#], Jingran Liu[2,#], Yunfei Hong[1], Yang Yang[1], Lou Li[1], Yilun Liu[2], Junkai Deng[1,*], Jefferson Zhe Liu[3,*]

[1]*State Key Laboratory for Mechanical Behavior of Materials, Xi'an Jiaotong University, Xi'an 710049, China*

[2]*State Key Laboratory for Strength and Vibration of Mechanical Structures, School of Aerospace Engineering, Xi'an Jiaotong University, Xi'an 710049, China*

[3]*Department of Mechanical Engineering, The University of Melbourne, Parkville, VIC 3010, Australia*

E-mail address: junkai.deng@mail.xjtu.edu.cn; zhe.liu@unimelb.edu.au.

[#]These authors contributed equally.

## Abstract

Polar topological structures in ferroelectric thin films have recently drawn significant interest due to their fascinating physical behaviors and promising applications in high-density nonvolatile memories. However, most polar topological patterns are only observed in the perovskites superlattices. Here, we report the discovery of the tunable ferroelectric polar topological defective structures designed and achieved by strain engineering in two-dimensional PbX (X=S, Se, and Te) materials using multiscale computational simulations. First, the first-principles calculations demonstrate the strain-induced recoverable ferroelectric phase transition in such 2D materials. The unique polar topological vortex pattern is then induced by applied mechanical indentation, evidenced by molecular dynamics simulations based on a developed deep-learning potential. According to the strain phase diagram and applied complex strain loadings, the diverse polar topological structures, including antivortex structure and flux-closure structure, are predicted to be emergent through the finite-element simulations. We conclude that strain engineering is promising to tailor various designed reversible polar topologies in ultra-flexible 2D materials, which provide excellent opportunities for next-generation nanoelectronics and sensor devices.

## Introduction

Topological objects with vortical configuration, such as skyrmion,[1–3] meron,[4] and other toroidal topological states,[5,6] come on stage with their intriguing properties in potential nonvolatile memory. Distinguished from the magnetic skyrmion of quasiparticles, the electric-polar topological structures with various vortex and polar skyrmion have been observed in recent years and developed the promising functionalities of these topological structures.[7–11] However, the discovered polar topological textures are limited in the ferroelectrics (FE) of perovskite structure, like the thin film of $(PbTiO_3)_n/(SrTiO_3)_n$ superlattice[12–14] and bubble domain of bulk $Bi_{0.5}Na_{0.5}TiO_3$.[15,16] Recently, the nanoscale topological materials, e.g., two-dimensional (2D) FE materials, are attracted significant interest as they satisfy the highly integrated and microminiaturized electronic element in the future.[17,18] Since 2D FE materials exhibit the intrinsic nanoscale thickness, natural ripple structure, and controllable polarization, they are reckoned to be good candidates for flexible nanoelectronics.[17–20] Although kinds of 2D FE materials have been predicted and even synthesized,[21] the exotic topological polar structures are rarely reported in 2D materials.

In this work, we report the discovery of tunable ferroelectric polar topological defective structures designed and achieved by strain engineering in two-dimensional PbX (X=S, Se, and Te) materials using multiscale computational simulations. The strain-induced paraelectric (PE) to FE transition in 2D lead monochalcogenides, PbX (i.e., PbS, PbSe, and PbTe), is demonstrated by density functional theory (DFT) based first-principles calculations. Different from other group-IV monochalcogenides (e.g., GeS, GeSe, SnS, or SnSe), whose ground state is black phosphorene-like structure (space group *Pnma*) with intrinsic ferroelectric polarization,[22] the ground state of PbX is a highly-symmetric paraelectric structure (space group *Cmcm*) without spontaneous ferroelectric polarization. Our first-principles calculation results reveal that the strain-induced PE to FE phase transition will occur for monolayer PbX material, leading to a puckered structure. The direction of polarization in these strain-induced FE states strongly depends on the applied external strain. Comprehensive calculations are then carried out to clarify the correlations between polarization states and complex strain fields.

Based on the developed deep-learning potential, the large-scale molecular dynamics (MD) simulations reveal that mechanical indentation will introduce complex inhomogeneous strain to monolayer PbX and obtain a vortex topological polar pattern. Three unique regimes of inhomogeneous strain fields are sequentially proposed through the strategy of strain engineering in such 2D materials. Using the finite element method (FEM) simulations, the local non-uniform strain fields and corresponding polarization patterns are predicted under uniform pressure, following the strain phase diagram of the first-principles calculation. Unexpectedly, three different kinds of polar topological patterns emerge by designing the shape of a hole and the layout of monolayer PbX on the substrate. It initially revealed the topological polar vortex in 2D materials and the coupling between mechanical loadings and polar vortex. The strategy to induce topological polar states of 2D materials by complex strain field broadens the way for the potential application of topological polar patterns.

## Results and Discussion

Among all reported 2D FE materials, group-IV monochalcogenides are a group of newly discovered FE materials whose heavily broken centrosymmetry and puckered structure endow them with large electric polarization.[22–24] In addition, recent studies demonstrated that the external strain and/or electric field-induced polarization switching and the multiferroic properties of group-IV monochalcogenides, implying their potential in electro-mechanical functionalization.[19,25–29] Monolayer PbX is in-plane isotropic with a lattice constant $a$ = 4.237, 4.400 and 4.638 angstrom for PbS, PbSe and PbTe, respectively. Unlike other group-IV monochalcogenides (e.g., GeS, GeSe, SnS, SnSe) with a noncentrosymmetric puckered structure (space group *Pnma*), the ground state of PbX shows a highly-symmetric structure in which the Pb and X atoms aligned in the *z* direction (space group *Cmcm*), indicating a non-polarized paraelectric phase (Fig. 1a). The phonon spectrum curve also demonstrates the structural stability of this PE structure even down to 0 K. Distinguished from traditional perovskite FE materials, whose high-temperature phase is PE state and transform into low-temperature FE states due to spontaneous symmetry breaking,[30–33] the PE state of PbX remains stable at low temperature.

Previous works have suggested an external strain could lead to a ferroelectric phase transition for thin films of the PE SrTiO$_3$ and KTaO$_3$.[33–36] Particularly, the SrTiO$_3$ films remain PE phase even down to 0 K. Its long-range ferroelectric order can be generated under a large enough epitaxial strain.[37] This well-known behavior inspires us to explore the possible ferroelectric phase by applying external strains in monolayer PbX. Unlike usually fragile perovskite structures, the superb mechanical deformability and flexibility of the 2D materials are expected to hold controllable deformed structures under large and complex strain fields.[38–40]

In order to examine the possible strain-induced phase transition in monolayer PbX, two kinds of applied strains were employed for investigation. Taking monolayer PbTe as an example, firstly, uniaxial strain along the $x$ direction was applied to the pristine PbTe (named as PE phase shown in Fig. 1a). With the applied tensile strain increased to a critical value of 3.5%, high symmetry PbTe crystals cannot hold its PE structure and the clear Pb-Te bonds reconstruction occurs to fit the stretch in $x$ direction. The generated phase shows a black phosphorene-like puckered structure with ferroelectric polarization along the $x$ direction, named as FE$_{[100]}$ phase (shown in Fig. 1b). Meanwhile, an in-plane shear strain is also applied to the PE phase. A similar ferroelectric structure with polarization along the diagonal direction, named as FE$_{[110]}$ phase (shown in Fig. 1c), was obtained when the shear strain was beyond 4%. The contour maps of electron localization function (ELF) for the three phases uncovered their structural difference. It is clear that the numbers of their hybrid orbitals are different, i.e., each Te atom shares 5, 3, and 4 covalent bonds with surrounding Pb atoms in PE, FE$_{[100]}$, and FE$_{[110]}$, respectively (Fig. 1d-f).

In the above stretch or shear process, the total energy keeps parabolically increasing till the critical strain has been reached. After that the energy follows a smoother parabolic curve, implying the PE-to-FE phase transition, as shown in the blue curves of Fig. 1g, h. The black and blue lines represent the energy change for the strained PE phase and strained FE phase, respectively. As the strain was released, the strained FE phase would recover back to the original PE phase, which shows the reversibility of the ferroelectric phase transition. Moreover, similar ferroelectric phase transitions also happened for PbS and PbSe with smaller critical tensile strain and shear strain (Supplementary Fig. 1). Their structural parameters and critical strain values are summarized in Table 1.

According to the thermodynamic theory of ferroelectric phase transition, spontaneous polarization is usually evaluated as the order parameter to distinguish the symmetry breaking in the phase transition.[41–43] For the displacive FE materials, the spontaneous polarization originates from the nonharmonic vibration of atoms, which leads to an atomic displacement relative to the original PE state, thus its order parameter can be estimated as the displacement distance.[44] Herein, in monolayer PbTe, the projected distance of Pb and Te atoms in the *x-y* plane is evaluated as the order parameter. As shown in Fig. 2a, the relative total energy ($\Delta E$) is considered as a function of the order parameter. When the applied tensile strain is lower than the critical strain (3.5% for PbTe), the total energy curve is parabolic increasing, and the state with zero order parameter (FE state) is the ground state. When the applied tensile strain is higher than the critical strain, two bistable FE states with non-zero order parameters arise, corresponding to the two equivalent states with non-zero polarization along the [-100] and [100] directions, respectively. The total energy curve and its derivative with respect to the strain (Supplementary Fig. 2) also imply the feature of continuous phase transition. In analogy with the phase transition in traditional FE perovskite materials, in which decreasing temperature induces the PE-to-FE phase transition, here, increasing external strain leads to the PE-to-FE phase transition in 2D FE PbX materials.

The soft mode theory of ferroelectric phase transitions revealed that there is a transverse optical lattice vibration mode whose frequency goes to zero as the transition is approached.[45–49] In the phase transition of monolayer PbX, the stability of sequent strained structures is also examined by phonon dispersion calculations (Supplementary Fig. 3-5). Taking monolayer PbTe as an example, the phonon dispersion spectra of the ground state shows no negative frequency indicating the dynamic stability. With the tensile strain increasing, the frequency of phonons gradually softens. Particularly, as shown in Fig.2b, a negative frequency (i.e., soft optical mode) at the gamma point rises when the applied strain is close to the critical point, implying structural instability. The soft optical modes related displacive instabilities have been assigned to be the origin of spontaneous symmetry breaking in the ferroelectric phase transition. Further analysis reveals that this soft vibration mode corresponds to the Te atom moving away from the Pb atoms in the *x* direction, as shown in Fig. 2c, which leads to the breaking of the centrosymmetry with polarization in the [100] direction. As a result, symmetry breaking leads to the phase transition from PE to FE phase. Similar phonon dispersion results are also obtained during the shear strain process, revealing the occurrence of spontaneous polarization in the [110] direction above critical shear strain.

Figure 3 shows the polarization change with increasing external strain. The Berry phase theory is employed to determine the polarization values.[50,51] The corresponding polarization with respect to the tensile strain and shear strain are shown in Fig. 3a and 3c, respectively. Before the critical strain, the polarization remains nearly zero, indicating the PE state. The phase transition from PE to FE accompanies an abrupt polarization change from zero to a non-zero value at the critical strain. Then, the polarization is continually enhanced with increasing strains showing a piezoelectric behavior. Meanwhile, the derivative of polarization with respect to strain is also evaluated, which represents the piezoelectric coefficient, as shown in Fig. 3 b, d. Notably, during the tension process, we found an enormous piezoelectric burst of $e_{11}$ (139.1, 144.3, and 139.2 ×$10^{-10}$ C/m for PbS, PbSe, and PbTe, respectively[52]) accompanying the phase transition, and finally converge to the stable piezoelectric coefficient $e_{11}$ about $2\times10^{-9}$ C/m. The convergent piezoelectric coefficients are comparable with other group-IV monochalcogenides, like monolayer SnS and SnSe ($e_{11}$ equals to 1.81 and 3.49×$10^{-9}$ C/m for SnS and SnSe, respectively). The specific value of the maximum and the stable $e_{11}$ are summarized in Table 2. A similar tendency appears in the shear process. The piezoelectric burst and the stable piezoelectric coefficient are also listed in Table 2. These enormous piezoelectric effects may provide promising applications in nanoscale sensors and energy harvesting devices.

The strain engineering of 2D materials is aimed to exploit mechanical strain to tune the electronic and photonic properties of 2D materials and to ultimately realize high-performance 2D-material-based functional devices.[36,39,40] Herein, the biaxial strain combined with shear strain, ranging from -7% to 7% in both *x* and *y* directions and 0 to 8% along the shear direction, was applied to the pristine phase of monolayer PbX to examine the PE-to-FE phase transitions. The polarization phase diagram of PbTe under biaxial strain and shear strain is illustrated in Fig. 4a. It shows there is a linear relationship between the spontaneous polarization and the fractional coordinates of the projected distance between two adjacent Pb and Te atoms. Thus, the phase diagram was plotted using the projector distance to present the spontaneous polarization which refers to the different phases. The grey areas with almost zero polarization refer to the initial PE phase. The other varicolored parts with a certain polarization refer to the FE phase. There is a clear boundary between the paraelectric phase and the ferroelectric phase. The corresponding directions of FE polarizations are demonstrated in Fig. 4b. When only biaxial strain is applied ($\gamma_{xy} = 0$), the polarization of the ferroelectric phase shows along the *x* direction when $\varepsilon_x > \varepsilon_y$, and vice versa. When $\varepsilon_x = \varepsilon_y$, the polarization shows along the diagonal direction. The other figures show the polarization maps with a certain shear strain $\gamma_{xy} = 0.02$, 0.04, and 0.06, respectively. When the shear strain is applied, the polarization direction falls in between [100] and [110] directions. The total polarization has components in both *x* and *y* directions, which is more complex under multiple strains of both biaxial stretch and in-plane shear. Meanwhile, the phase boundary shifts towards negative in both *x* and *y* axes when the shear strain increases. It reveals the change in the critical strain of ferroelectric phase transition and might explore the functionality of designable ferroelectricity. Especially, the polarization phase diagrams provide a guideline to tune the FE polarization and further design the polarization patterns for monolayer PbX under mechanical loadings.

As the limitation in first-principles calculations, it is difficult to carry out the mechanical loading for the large-scale models of PbX. Thus, we develop a deep-learning potential to perform large-scale molecular dynamics (MD) simulations to investigate the polarization distribution of a monolayer PbX under strains. As shown in Fig. 5a, a spherical tip with a radius of 3.5 nm is employed to introduce mechanical indentation to a monolayer PbTe supercell with 60 × 60 units (i.e. about 28×28 nm). The applied strain field is inhomogeneous in this manner and the local strain varies in each unit-cell. When the depth of indentation reaches 3.6 nm, the local strain concentration is beyond the critical strain for FE phase transition. Consequently, an interesting vortical polarization pattern arises in the center of the indentation. As shown in Fig. 5b, four domains are divided by the diagonal lines, and the net polarization of each domain nearly align along the *x* or *y* axis. A domain wall of 90 degrees concatenates them to reduce the electrostatic energy and strain energy, which are brought by the depolarization fields and the strained lattice. It demonstrates that the inhomogeneous strains lead to unusual topological polar structures.

Although each unit-cell of the indented membrane is no longer located on the original *x-y* plane, the local strains are determined with respect to the pristine unit-cell. The distortion of each unit-cell can be divided into three components, including the relative change of lattice parameters *a* and *b*, and the shear deformation of the unit-cell in the local tangent plane, as demonstrated in Fig. 5c-e. The distribution of these distortion components is also approximately distinguished along the diagonal lines. It is found that the local strain distributions are closely related to the polarization pattern. It inspired us to further explore more topological polar defective patterns on monolayer PbX under complex nonuniform strain conditions.

The natural flexibility and considerable strength of 2D PbX materials endow them with potential application in electro-mechanical devices. Motivated by recent state-of-art strain engineering technology of 2D materials,[53] we designed three models to investigate the possible strain/polar distribution of monolayer PbX under complex strain fields in device scale (~um). Therefore, the FEM simulations rather than MD simulations are employed to accomplish this task. The basic design is to put the 2D monolayer membrane onto a specific artificial substrate that has a hole in the middle and kept the boundary clamped. Then, an external uniform pressure is applied through the hole to form a bubble-like structure. As the clamped edge of the hole, the membrane was pressed to generate an inhomogeneous strain field. The differences between these models are the shape of the hole and the direction to arrange the 2D film on the substrate. As shown in Fig. 6a, an around and isotropous hole model with a diameter of 200 nm under a downward blowing pressure was designed. Similarly, in Fig. 6b, a square hole with an edge length of 200 nm is proposed, and the downward pressure is subsequently applied. In this model, the direction of the lattice of the 2D material is deliberately arranged along the edge of the hole. While in Fig. 6c, the direction of the lattice is arranged to keep an angle of 45 degrees with the hole edge, i.e. along the diagonal direction. The required mechanical parameters of the PE and FE phase are obtained based on DFT calculations, which are elaborated in the Supplementary Table 1. The elastic constant, $C_{11}$, of the FE phase is only 6.228 Nm$^{-1}$, far less than that of the PE phase (i.e., 38.064 Nm$^{-1}$), indicating the softness along the armchair direction. Meanwhile, their elastic constant in $y$ direction, $C_{22}$, is close, indicating the relative hardness in this direction.

Firstly, the strain fields of these 2D PbTe membranes are investigated without the consideration of phase transition. Supplementary Fig. 6 demonstrated the distribution of $\varepsilon_x$, $\varepsilon_y$, and $\gamma_{xy}$ in these three models, respectively. It reveals the strain fields are closely dependent on the shape of the hole and the alignment direction of the membrane. According to the above results, in principle, ferroelectric phase transitions will occur when the strain is beyond the phase boundary in the phase diagram (Fig. 4). Thus, the FE phase is possible to firstly nucleate at the area of strain concentration. Meanwhile, the continuous strain filed evolution demonstrates the validity of the mechanical loadings (Supplementary Fig. 7).

To investigate the possible polarization pattern precisely, we considered the phase transition and the mechanical difference between the FE and PE phase in the loading process of the above three models. In the calculation of strain fields, the strain criterion is implemented additionally, i.e., when the strain of the current unit is within the PE phase regime, the mechanical parameters of the PE phase are applied. While the strain is beyond the phase boundary to reach the FE phase regime, the mechanical parameters of the FE phase are applied instead. The calculated strain fields are demonstrated in Supplementary Fig. 8. There have apparent differences compared with the former strain fields without the consideration of phase transition (Supplementary Fig. 7). Sharp boundaries arise and the area of strain concentration also changed in new strain fields. At the beginning of pressure loading, the strain in each unit is relatively small and lower than the critical strain of phase transition. The strain fields vary the same with the condition in Supplementary Fig. 6. However, the FE phase emerges with further increasing loadings.

In model I, although the strain concentrates at the center area at the beginning, the FE phase then nucleate in the vicinity of diagonal lines at critical strain condition, and their polarizations are arranged along the principal axis (*x* and *y* directions). Subsequently, the FE phase grows with the inward extending of strains. The evolution process is shown in Supplementary Fig. 9a. When the load pressure is 10 MPa, the FE phase occurs near the diagonal lines. After the FE phase nucleates from these parts, polarizations here are further enhanced with increasing load. The polar structure evolution was observed until the load pressure of 14 MPa, and the final polarization map is plotted in Fig. 6d. An anti-vortical topological polar structure is obtained. The background color indicates the magnitude of polarization. It finds that the closer to the diagonal lines, the larger polarization. In model II, however, the strain firstly concentrates at the edges. FE phase nucleates at the edges and grows toward the inner regime. As shown in Supplementary Fig. 9b, the polar structure evolution was finally completed at the load pressure of 8 MPa. Fig. 6e. demonstrated another anti-vortical topological pattern observed as the final polar distribution. The relatively uniform background color implies the possible homogeneous nucleation in this model. At last, in model III, the gradient of strain is rather big and the strain concentrates at the center heavily. As shown in Supplementary Fig. 9c, the FE phase arise at the center and grows rapidly. The FE domains saturate only at the load pressure of 4.5 MPa. Fig. 6f demonstrates the final distribution of polarization is a flux-closure topological pattern, which is a similar structure observed in ferromagnetic materials and ferroelectric perovskite materials.[6,54,55]

Our predicted topological polar structures are quite diverse depending on the shape of the hole and the arranged manner of the membrane. More complicated shapes, i.e., polygonal holes and sectorial holes, could be carefully designed and bring different effects to the topological polar structures.[40] Moreover, as mentioned above, the present FE phase transition is reversible for monolayer PbX. Once the strain field is released, the FE phase turns back to the pristine PE phase. It suggests that the triggered topological defective patterns can be controllable and removable by external pressure. Furthermore, other 2D FE materials with intrinsic FE polarizations also have great potential to apply complex strain loadings and corresponding deformations. It is thus reasonable to predict the various complicated topological polar structures achievable by strain engineering in other easy-deformable 2D FE materials. As studied in recent works, the exotic topological defects in ferroelectric perovskite materials, like flux-closure domains, skyrmion lattice, and polar meron, present novel properties under mechanical, optical, and electric fields.[3,8] The discovery of similar topological polar structures in 2D materials developed the knowledge of polar skyrmion fields and revealed the potential application of those various tunable topological patterns of 2D FE materials in future high-density storage.[11]

## Conclusion

In this work, multiscale simulations were utilized to discover the strain-induced PE-to-FE phase transition and design the artificial dipole topological patterns in 2D Pb-monochalcogenide. The intrinsic high-symmetry PE structure of PbX endows them strain dependent polarizations based on the found strain-induced PE-to-FE mechanism. The mechanical indentation can introduce inhomogeneous strain fields to PbTe membrane, and lead to unique FE polarization patterns in local regions. Further FEM simulation reveals that the types of polarization patterns are closely dependent on the hole shape of the substrate and the layout of the membrane. The result enriched the knowledge of 2D polar topological patterns and broaden the strain engineering of 2D materials in the devices of electro-mechanical coupling.

## Methods

**DFT calculations.** The first-principles calculations in this work were performed based on the density-functional theory (DFT), as implemented in the Vienna Ab initio Simulation Package (VASP).[56–59] Projector augments wave pseudopotentials with the Perdew-Burke-Ernzerhof (PBE) exchange-correlation and generalized gradient approximation (GGA) were employed.[60–62] For all calculations, the plane-wave cutoff was set to be 600 eV. A Monkhorst-Pack gamma-centered k-points grid of dimensions 25×25×1 was adopted for the unit cell.[63] The cell height in the direction normal to the surface plane was fixed as 20 Å to avoid the interactions between adjacent layers. In all cases, the atoms were fully relaxed in all directions until the force on each atom was less than 0.001 eV/Å, ensuring the accuracy for the optimization of structure. The electric polarization was directly computed via "Berry-phase" theory.[50]

**Machine learning potential.** The MD simulation with machine learning potential was performed to get an insight on the atomic picture of the PbTe under strain and finite temperature. With the help of this, the evolution of the atoms during the MD simulation can be calculated with a considerable accuracy, but at a much less computational cost.[64] The machine learning potential is constructed by the DeepMD packages[65] based on the dataset provided by DFT calculations, as demonstrated in the flow chart of Supplementary Figure 10. The configurations under different strain and temperature (10 ~ 400K) were sampled by the short AIMD simulations picked up every 10 steps. The whole dataset consists of 100K different labeled samples. AIMD calculation of PbTe were performed using VASP.

In the DeepMD model, the potential energy E of a configuration is assumed to be a sum of each atomic energy $E_i$ of atom $i$, which is fitting from a descriptor $D_i$ through an embedding network. The descriptor $D_i$ characterizes the local environment of atom $i$ within a cutoff radius $R_c$. Here, the $R_c$ is set to 7.5 Å. The maximum number of atoms within the $R_c$ is set to 95 for Pb, 95 for Te, respectively. The translational, rotational, and permutational symmetry of the $D_i$ are preserved by an embedding network. The smooth edition of the DeepMD model was employed to remove the discontinuity introduced by the cutoff radius. The inner cutoff where the smooth begins was set to 7.2 Å.[66] The sizes of the embedding and fitting networks are (25, 50, 100) and (240, 240, 240), respectively. The loss function is defined as

$$L(p_\varepsilon, p_f, p_\xi) = p_\varepsilon \Delta \epsilon^2 + \frac{p_f}{3N} \sum_i |\Delta F_i|^2 + \frac{p_\xi}{9} \|\Delta \xi\|^2$$

where $\Delta \epsilon$, $\Delta F_i$, and $\Delta \xi$ represent the energy, force, and virial tensor difference between the DeepMD model prediction and the training dataset (from DFT calculations). The $p_\varepsilon$, $p_f$, and $p_\xi$ are weight coefficients of energy, force, and virial tensor. The $p_\varepsilon$ increase from 0.02 to 1, and $p_f$ decreases from 1000 to 1 during the training procedure. The virial tensor was discarded in this study, thus the $p_\xi$ is set to 0. The DeepMD models are trained for 6 000 000 steps. Subsequently, various tests have been made to testify the accuracy of our machine learning potential.

The obtained machine learning potential performs well at larger scale and can reproduce the phase transition results of DFT calculations, as shown in Supplementary Figure 11. A detailed verification of this potential is elaborated in another work.[67]

**MD simulation**

With the well-testified machine learning potential, equilibrium MD simulations were performed using LAMMPS packages in a temperature of 10K.[68] The 60 × 60 2D PbTe supercell (14400 atoms) was used and the duration of the MD simulation was 90 ps with a 1 fs time step. The atomic velocities and positions were collected every 1 ps under the NVT ensemble for the calculation of the offset ($\vec{R}$) of nearest Pb and Te atoms. This offset is taken as the polarization of the 2D PbTe in the specific position and then projected to the XOY plane to give a better visual presentation.

**FEM methods**

The unique strain fields of 2D PbTe at macro-scale were designed by FEM simulations. The 2D PbTe is taken as an elastic anisotropic membrane with different elastic constants before and after the phase transition in our simulation. The parameters used in our calculation are listed in Supplementary Table 1. During the indentation, it is fixed on the matrix by restricting all the degree of freedom around the edges and the indenters and holes with different shape were adopted. Subsequently, the polarization direction was decided according to polarization-strain diagram obtained by DFT calculations.

## Table 1:

Table1: The lattice parameters and critical strain of phase transition for monolayer PbX

| Materials | $a$ (Å) | $\varepsilon_x$ or $\varepsilon_y$ | $\gamma_{xy}$ |
|---|---|---|---|
| PbS | 4.237 | 1.6 % | 2.1 % |
| PbSe | 4.400 | 2.2 % | 2.4 % |
| PbTe | 4.638 | 3.5 % | 3.6 % |

## Table 2:

Table2: The piezoelectric coefficients for monolayer PbX under mechanical loadings

| Materials | Stretch | | Shear | |
|---|---|---|---|---|
| | Max $e_{11}$ ($10^{-10}$C/m) | Final $e_{11}$ ($10^{-10}$C/m) | Max $e_{16}$ ($10^{-10}$C/m) | Final $e_{16}$ ($10^{-10}$C/m) |
| PbS | 139.1 | 19.0 | 115.3 | 9.3 |
| PbSe | 144.3 | 19.1 | 101.9 | 6.3 |
| PbTe | 139.2 | 20.4 | 122.8 | 9.4 |

Figure 1:

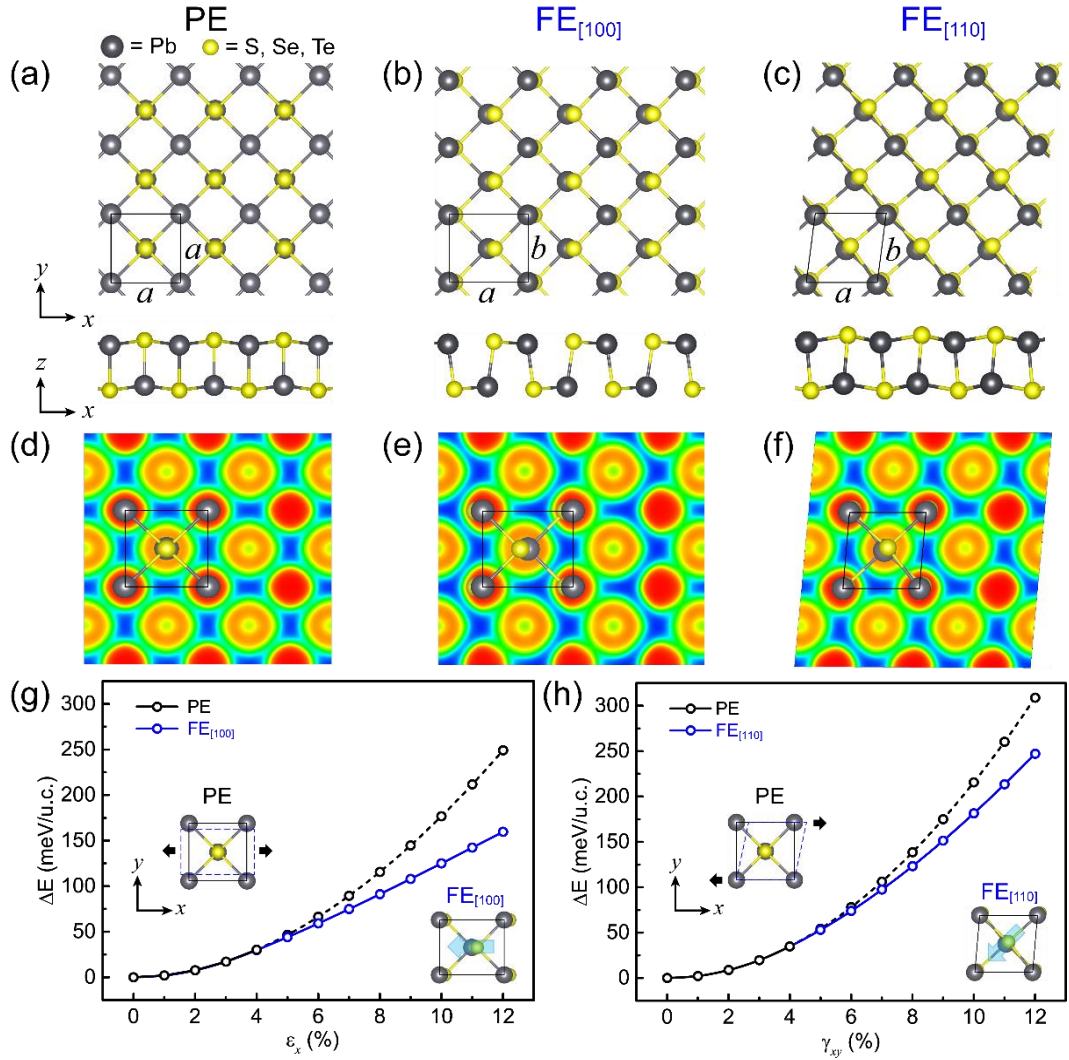

Figure 1. The PE, FE$_{[100]}$, and FE$_{[110]}$ phases for monolayer PbX and the energy variation under mechanical loadings. (a). The atomic structures of the PE, FE$_{[100]}$, and FE$_{[110]}$ phases for monolayer PbX. (d-f) The corresponding electron localization function (ELF) for the three phases. There are five, three, and four atomic bonds for each chalcogenide atom in PE, FE$_{[100]}$, and FE$_{[110]}$ phases, respectively. (g-h) The energy variation of monolayer PbTe under uniaxial stretch and in-plane shear. The parabolic energy curve turns into a smoother curve when the applied stretch and shear strain is beyond 3.5% and 3.6%, respectively.

Figure 2:

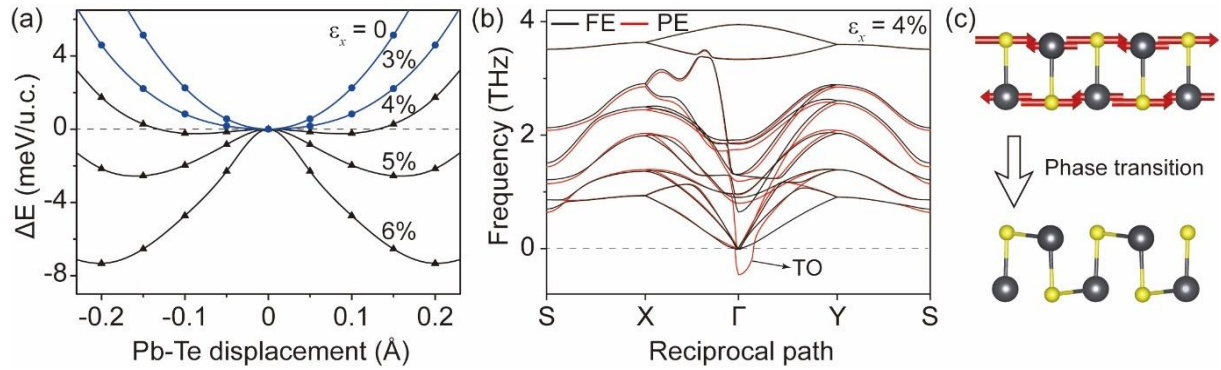

Figure 2. The strain-induced PE-to-FE phase transition originates from the transverse optical lattice vibration mode. (a) The relative total energy as a function of Pb-Te displacement under different stretch strains. Below the critical strain of 3.5%, the PE phase is the ground state. Otherwise, two bistable FE phases are more stable than the PE phase when the strain is larger than the critical value. (b) The phonon dispersion spectra of the PE and FE phases for monolayer PbTe under the stretch strain of 4%. The black lines of the FE phase indicate its stability, while the red lines of the PE phase indicate the soft transverse optical (TO) mode at Γ point, corresponding to its instability. (c) The soft TO mode is related to corresponding atomic vibration, i.e., the inverse vibration of Pb and Te atoms along $x$ direction.

Figure 3:

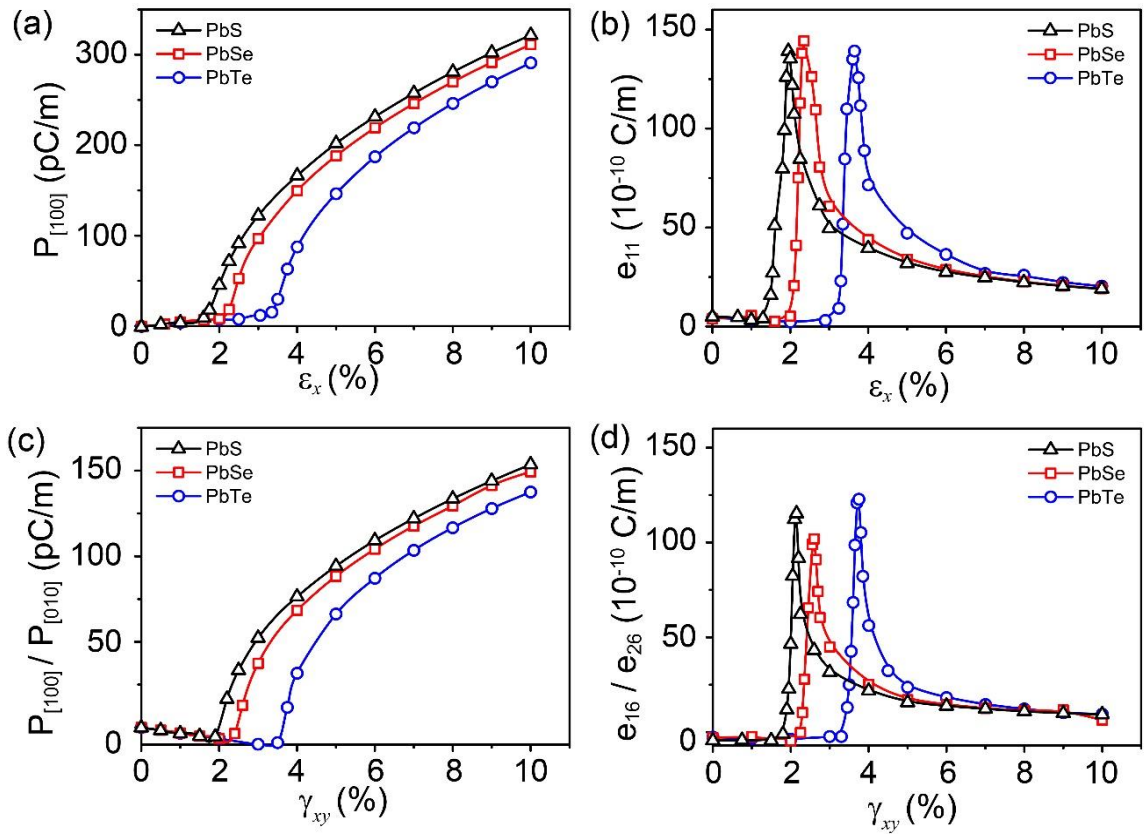

Figure 3. The ferroelectric polarization and corresponding piezoelectric coefficient as a function of applied strain. (a) The FE polarization changes under uniaxial strain. (b) The piezoelectric coefficient $e_{11}$ changes under uniaxial strain. (c) The FE polarization changes along [100] / [010] direction under shear strain. (d) The piezoelectric coefficient $e_{16}$ / $e_{26}$ changes under shear strain.

Figure 4:

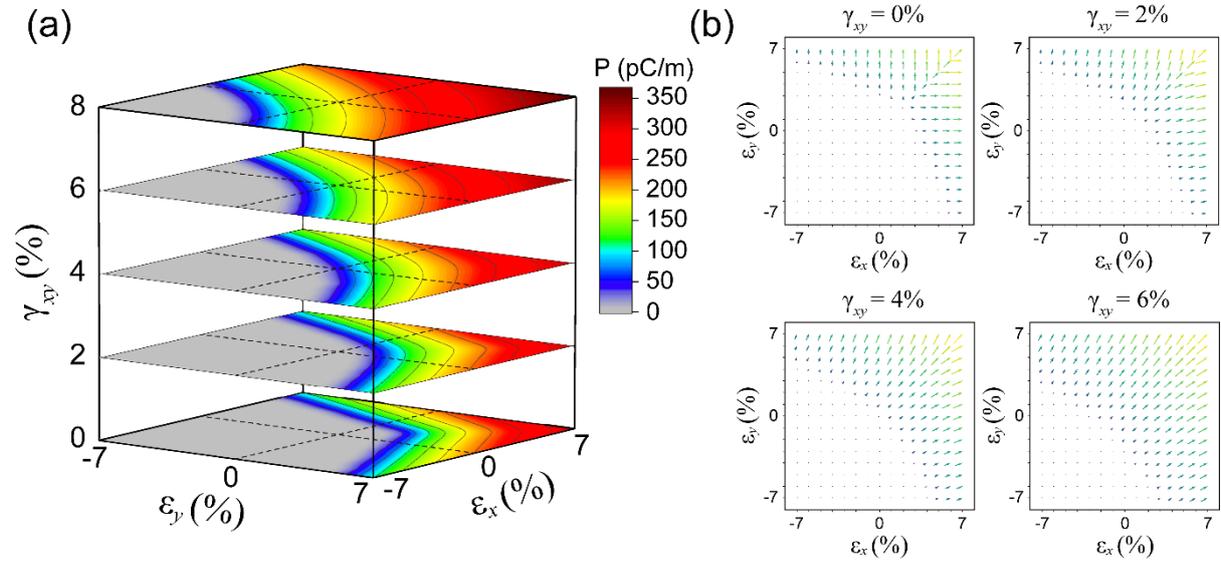

Figure 4. Polarization diagram of monolayer PbTe. (a) The whole polarization diagram under complex strain fields combined the in-plane biaxial strain and the shear strain. The boundary between grey and other colored area represents the phase boundary between PE and FE phases. (b) The polarization map under in-plain biaxial strain combined with the shear strain of 0%, 2%, 4%, and 6%, respectively. The arrows represent the polarization values and directions at a specific strain condition.

Figure 5:

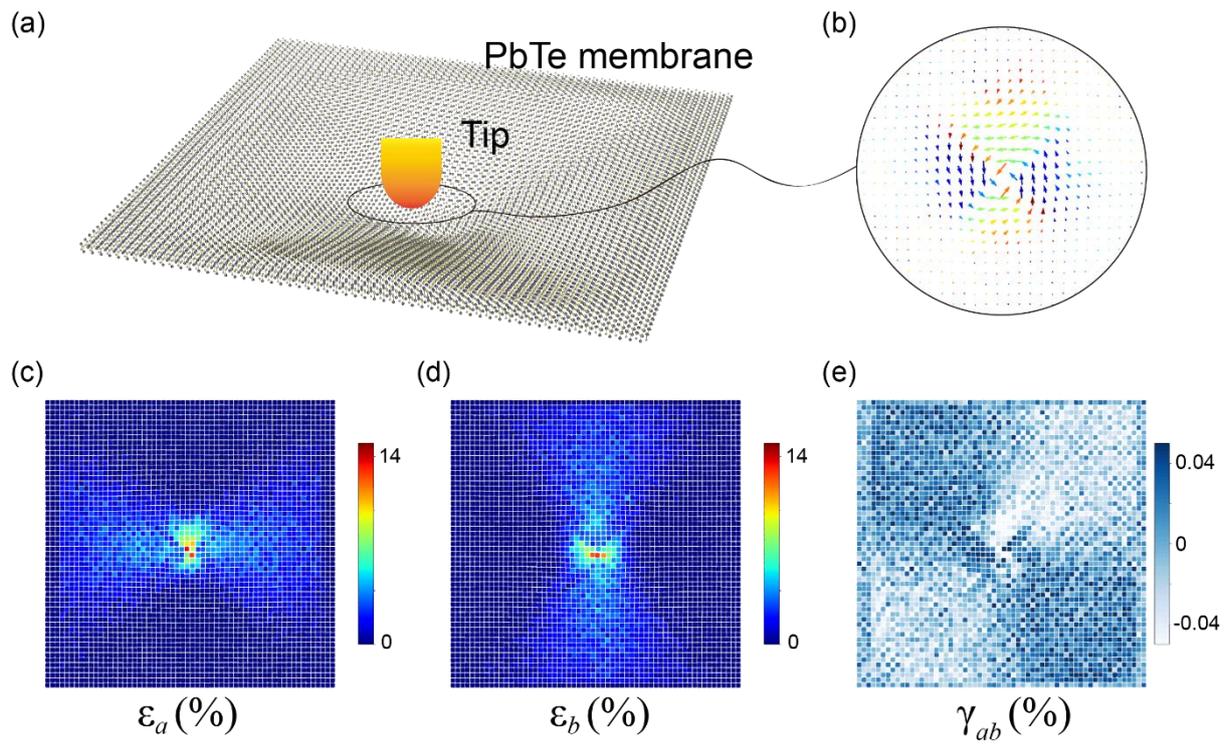

Figure 5. Atomic configuration of monolayer PbTe under mechanical indentation and corresponding polar distribution as well as related lattice strain components distribution of MD simulation. (a) The mechanical indentation of monolayer PbTe by a tip. (b) The polar vortex structure was observed in the center of monolayer PbTe. Each arrow represents the net polarization of a unit-cell. (c)-(e) The distribution of three components of lattice distortion under mechanical indentation. Each block unit represents the local lattice strain of a unit-cell, including the relative change of lattice parameters *a* and *b*, and the shear distortion of the unit-cell in the local tangent plane. The three components are approximately divided along the diagonal direction of the *x-y* projective plane. It implies the possible polarization pattern is closely dependent on the inhomogeneous strain distribution.

Figure 6:

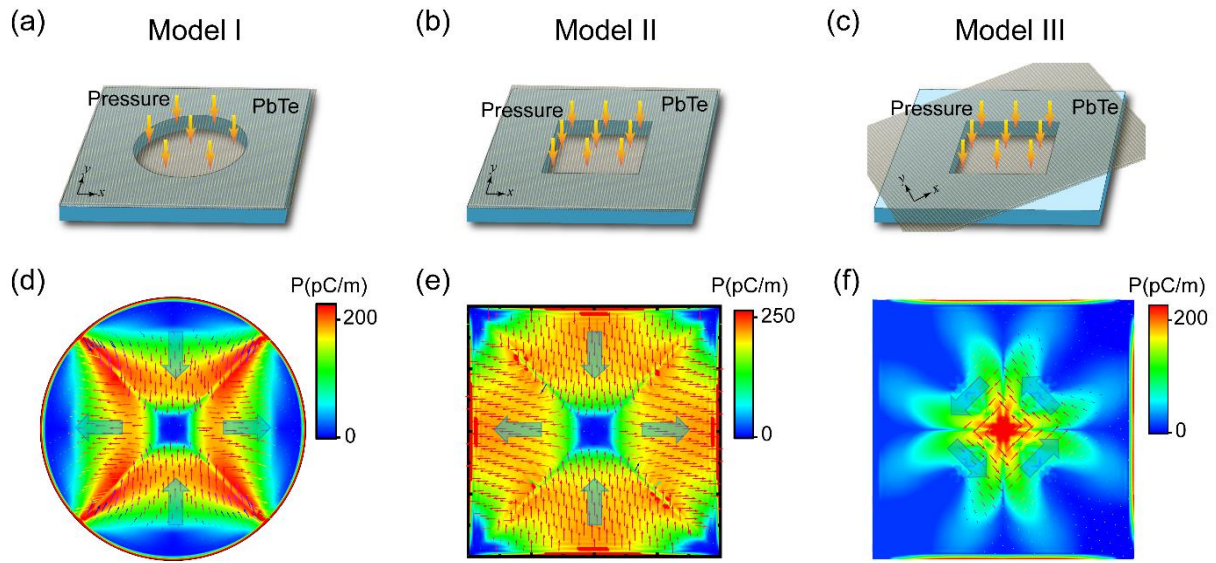

Figure 6. Three designed models to apply strains on the monolayer PbTe membrane and corresponding obtained topological polar structures. (a) The model I: monolayer PbTe membrane onto a substrate with a circle hole. A "blowing" downward uniform pressure is applied to the membrane. (b) The model II: monolayer PbTe membrane onto a substrate with a square hole. The direction of the lattice is arranged along the hole edges. (c) The model III: the monolayer PbTe rotates 45 degrees based on model II. (d) The cubic anti-vortex polarization pattern is predicted at the load pressure of 14 MPa on the Model I. (e) The round-shape anti-vortex polarization pattern is predicted at the load pressure of 8 MPa on the Model II. (f) The flux-closure polarization pattern is predicted at the load pressure of 4.5 MPa on Model III.

Supplementary Information for

# Tunable ferroelectric topological defects on 2D topological surfaces: strain engineering skyrmion-like polar structures in 2D materials

Bo Xu[1,#], Zhanpeng Gong[1,#], Jingran Liu[2,#], Yunfei Hong[1], Yang Yang[1], Lou Li[1], Yilun Liu[2], Junkai Deng[1,*], Jefferson Zhe Liu[3,*]

[1]*State Key Laboratory for Mechanical Behavior of Materials, Xi'an Jiaotong University, Xi'an 710049, China*

[2]*State Key Laboratory for Strength and Vibration of Mechanical Structures, School of Aerospace Engineering, Xi'an Jiaotong University, Xi'an 710049, China*

[3]*Department of Mechanical Engineering, The University of Melbourne, Parkville, VIC 3010, Australia*

E-mail address: junkai.deng@mail.xjtu.edu.cn; zhe.liu@unimelb.edu.au.

[#]These authors contributed equally.

Supplementary Table 1:

Table1: The mechanical parameters of PE and FE phases of monolayer PbTe

| Phase | $a$ (Å) | $b$ (Å) | $C_{11}$ (N/m) | $C_{22}$ (N/m) | $C_{12}$ (N/m) | $C_{44}$ (N/m) | $\upsilon_{12}$ | $\upsilon_{21}$ |
|---|---|---|---|---|---|---|---|---|
| PE | 4.638 | 4.638 | 38.064 | 38.064 | 20.554 | 31.990 | 0.54 | 0.54 |
| FE | 4.670 | 4.580 | 6.228 | 30.476 | 8.783 | 18.983 | 0.29 | 1.41 |

Supplementary Figure 1:

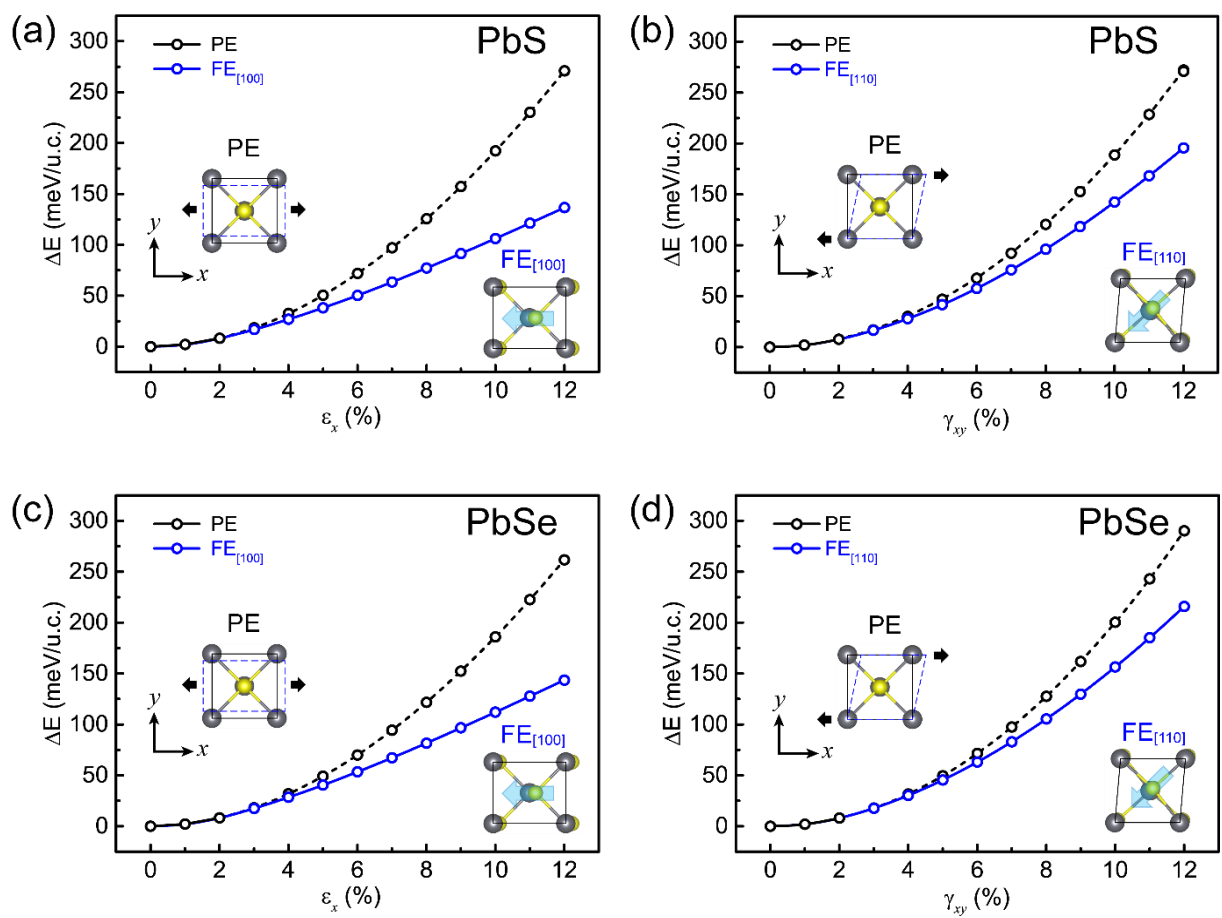

Supplementary Figure 1. The energy variation of monolayer (a-b) PbS and (c-d) PbSe under uniaxial stretch and in-plane shear. The parabolic energy curve turns into a smoother curve when the PE-to-FE phase transition occurs.

Supplementary Figure 2:

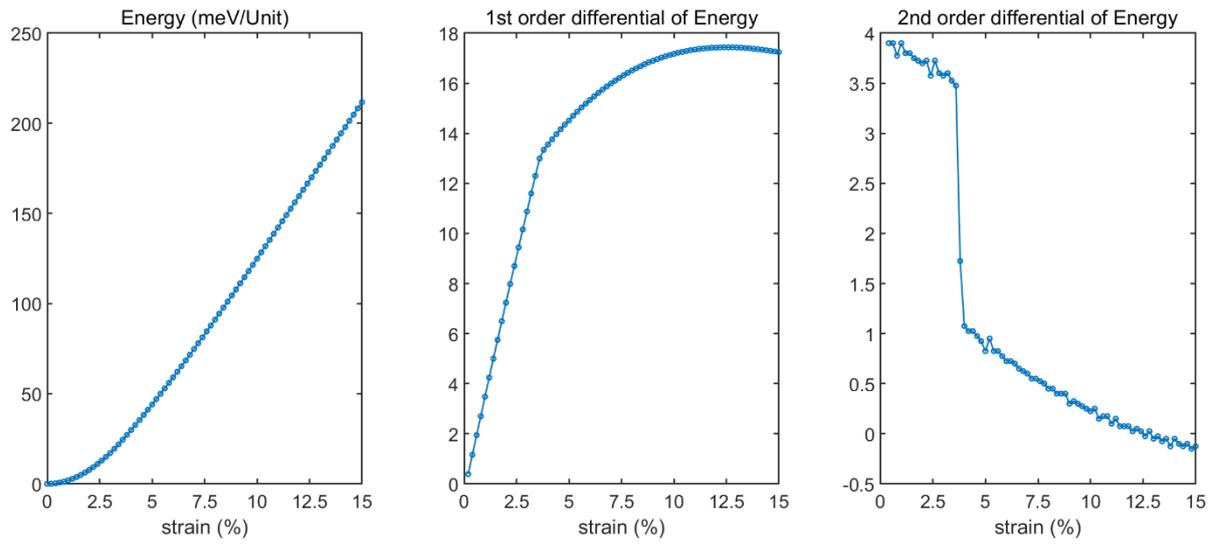

Supplementary Figure 2. The (a) energy variation of monolayer PbTe under uniaxial stretch, and corresponding (b) first-order and (c) second-order differential of energy with respect to the strain. The first-order differential of energy is still continuous when phase transition happens, while the second-order differential is discontinued at the critical point of phase transition.

Supplementary Figure 3:

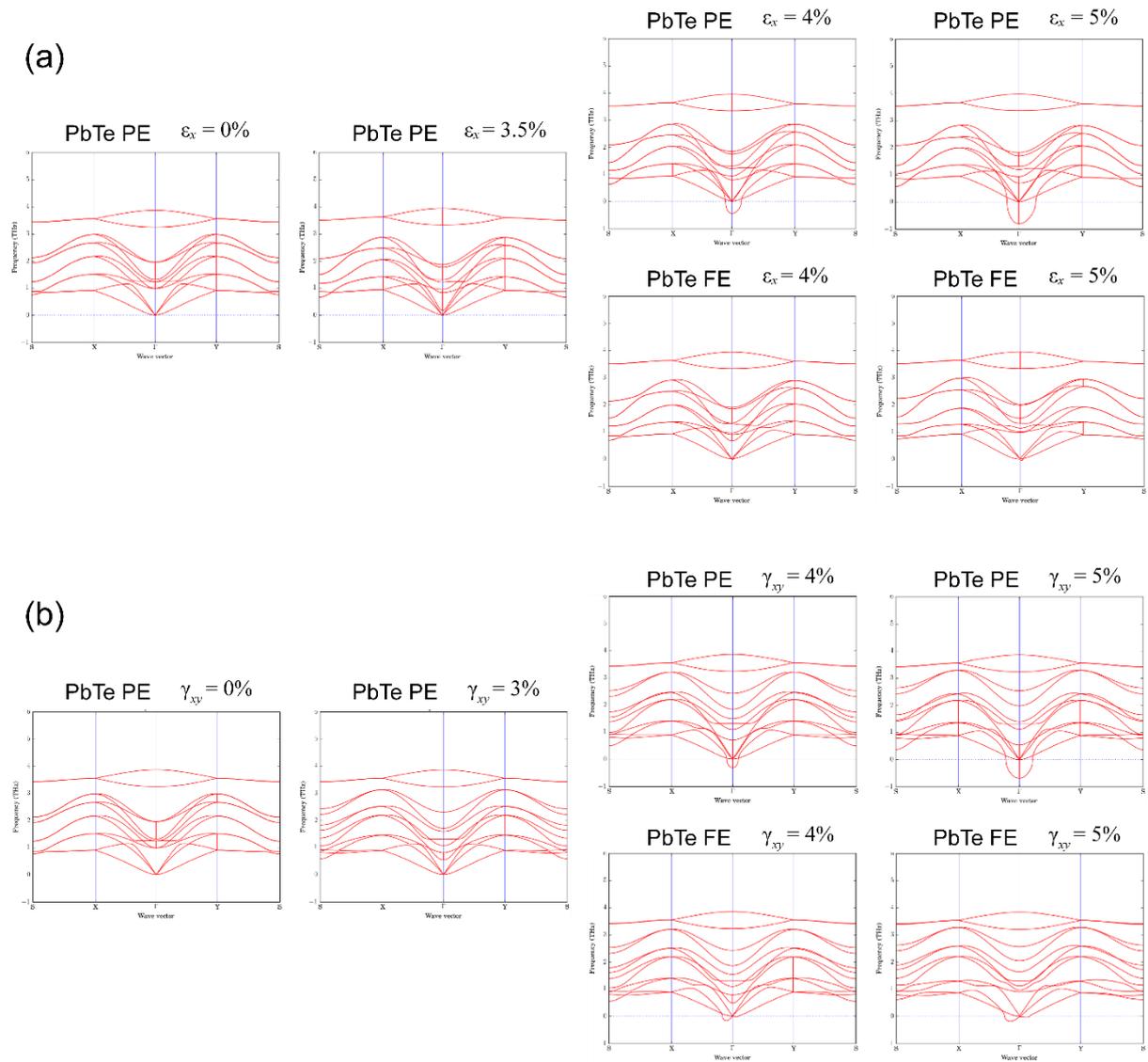

Supplementary Figure 3. The phonon dispersions of strained monolayer PbTe under different (a) uniaxial stretch and (b) in-plane shear. When the applied strain is beyond the critical point of phase transition, the negative frequency of modes arises in the phonon dispersion of the strained PE structure, indicating the instability of the PE structure under this strain condition. Whereas no soft mode appears in the phonon dispersion of FE structure under this corresponding strain, indicating its stability.

Supplementary Figure 4:

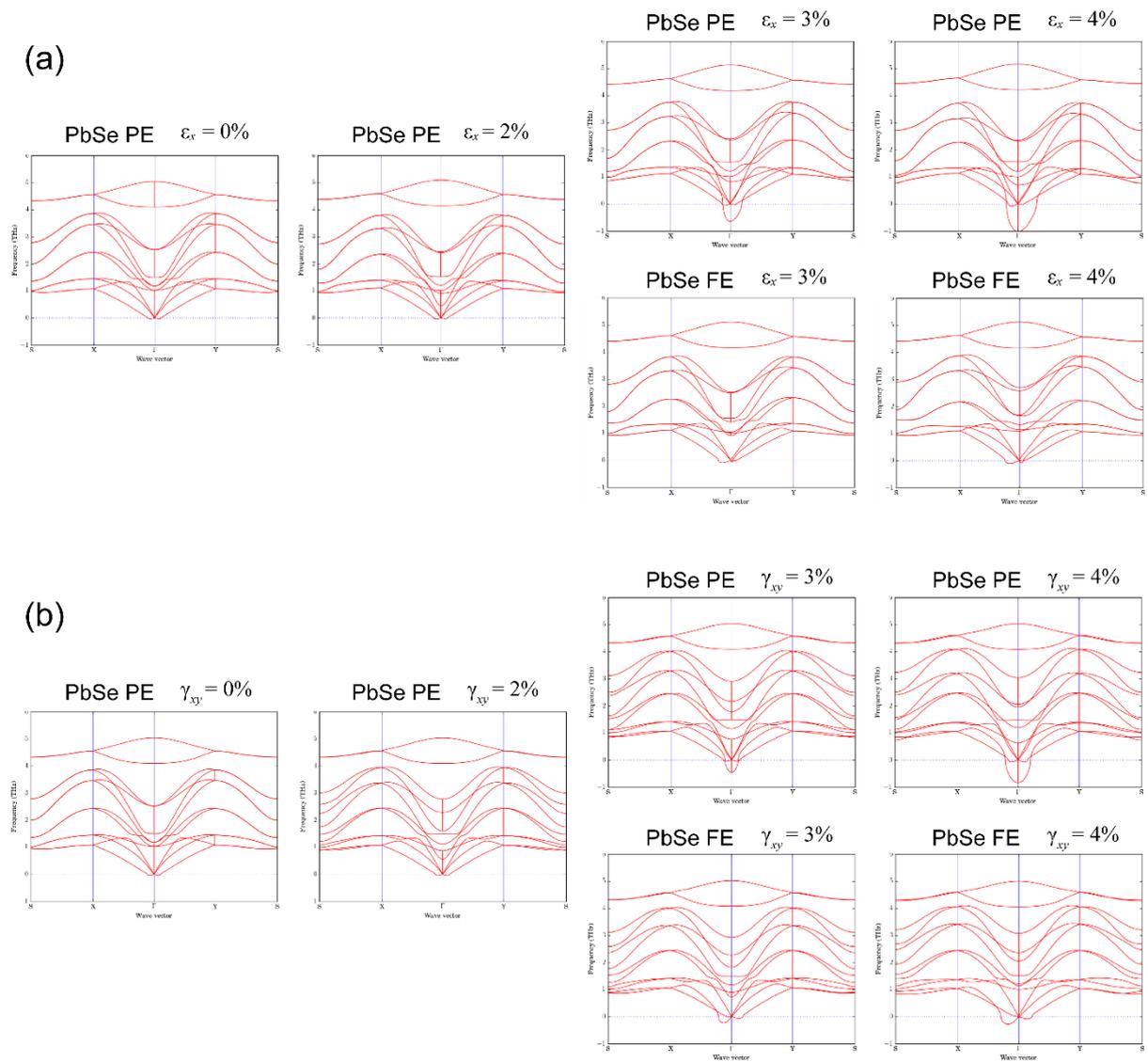

Supplementary Figure 4. The phonon dispersions of strained monolayer PbSe under different (a) uniaxial stretch and (b) in-plane shear. When the applied strain is beyond the critical point of phase transition, the negative frequency of modes arises in the phonon dispersion of the strained PE structure, indicating the instability of the PE structure under this strain condition. Whereas no soft mode appears in the phonon dispersion of FE structure under this corresponding strain, indicating its stability.

## Supplementary Figure 5:

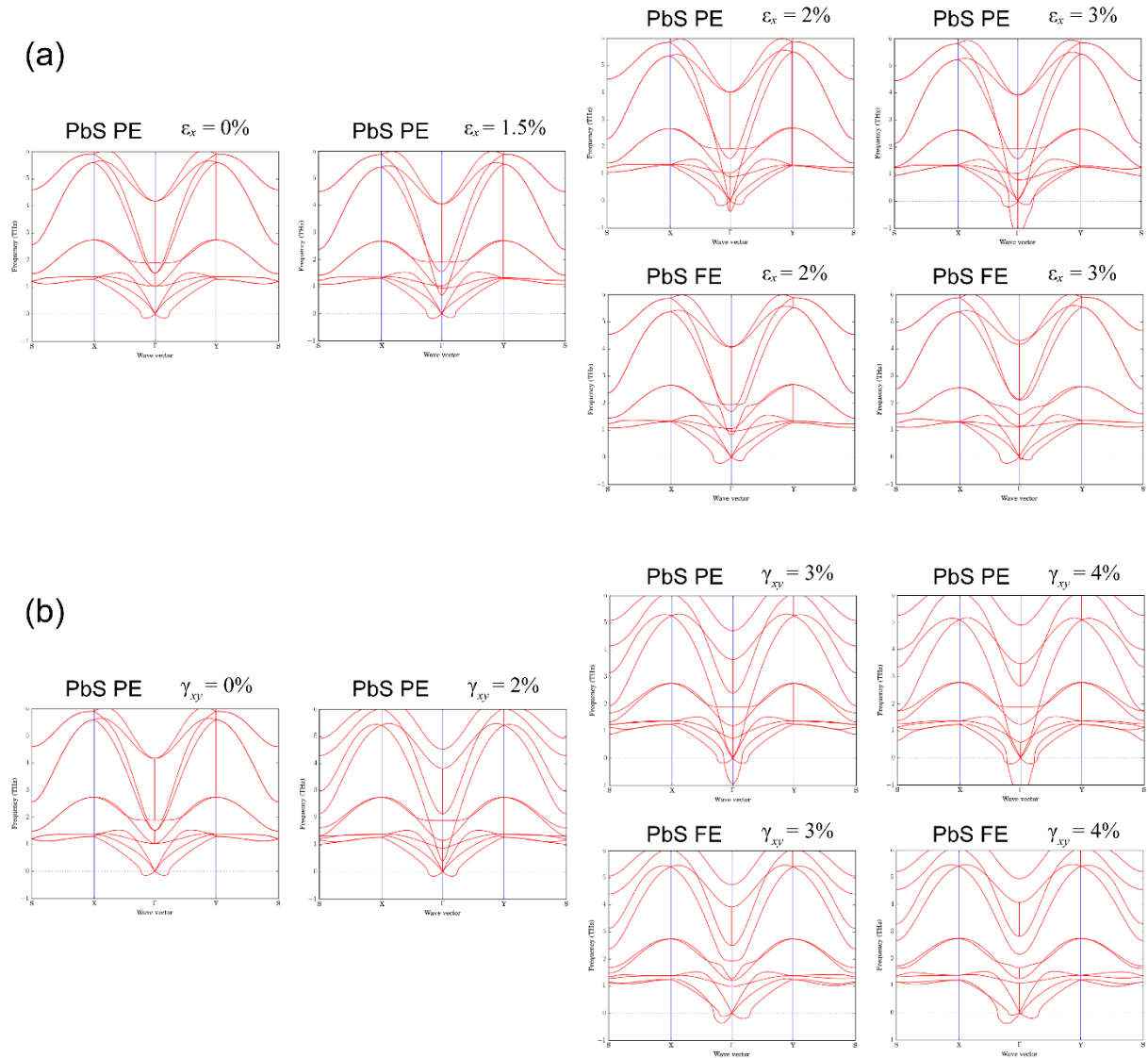

Supplementary Figure 5. The phonon dispersions of strained monolayer PbS under different (a) uniaxial stretch and (b) in-plane shear. When the applied strain is beyond the critical point of phase transition, the negative frequency of modes arises in the phonon dispersion of the strained PE structure, indicating the instability of the PE structure under this strain condition. Whereas no soft mode appears in the phonon dispersion of FE structure under this corresponding strain, indicating its stability.

Supplementary Figure 6:

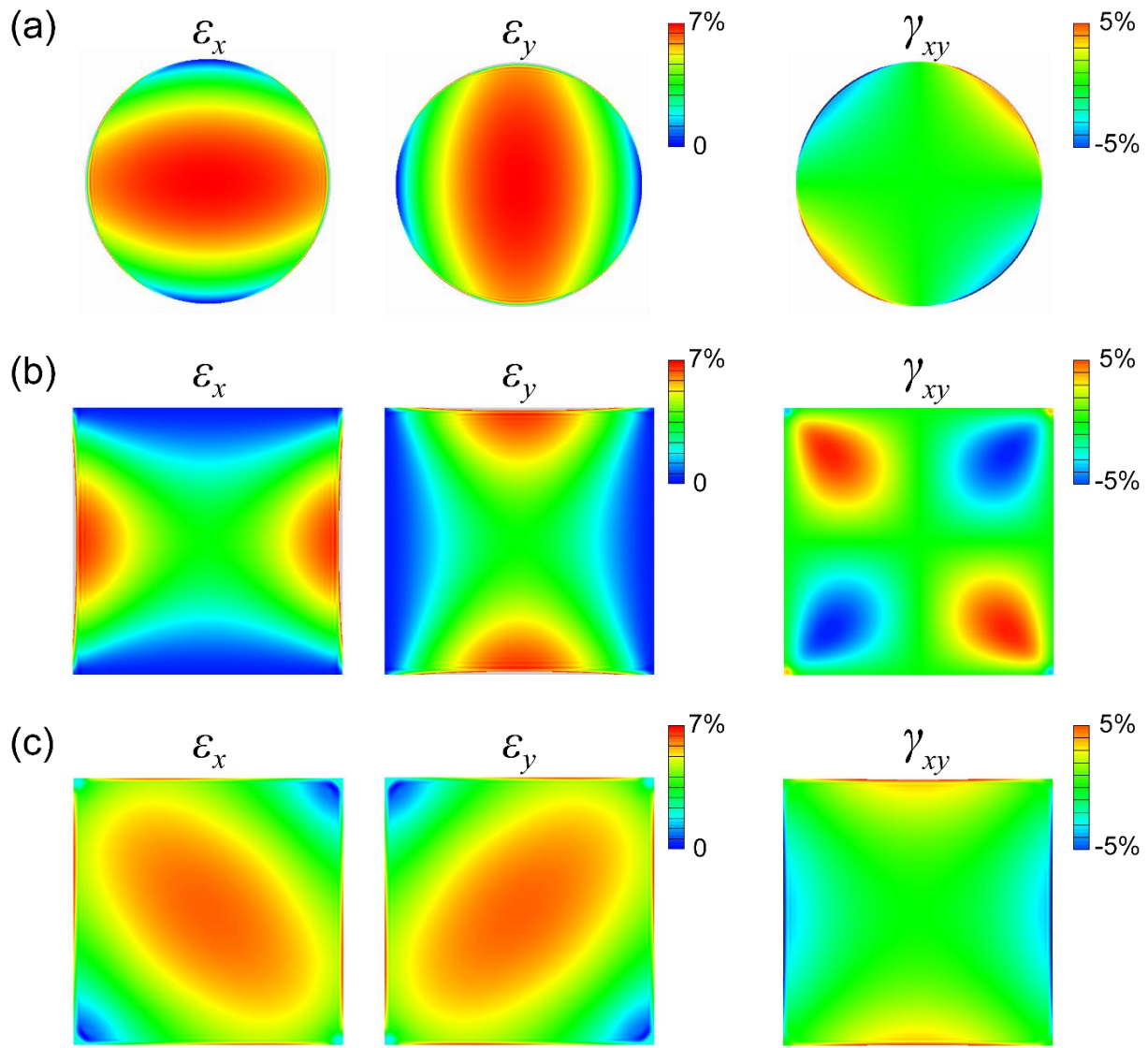

Supplementary Figure 6. The distribution of strain components along the *x*, *y* direction and shear strain in the *x-y* plane of monolayer PbTe membrane under three designed loading models. The results here have not taken phase transition into consideration, which means the input parameters for FEM simulation here are only the mechanical properties of PE structure, for comparing with the results in Supplementary Figure 8.

Supplementary Figure 7:

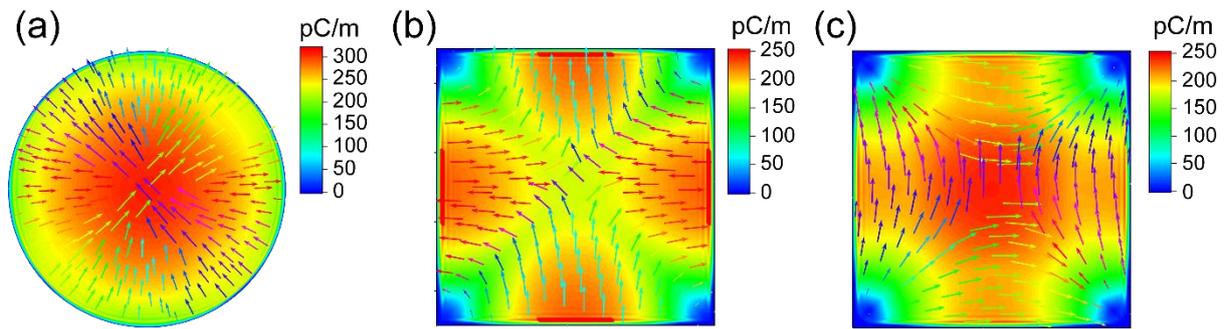

Supplementary Figure 7. The final distribution of polarization and total strain of three loading models without the consideration of phase transition. Compared with the distribution in Figure 6 of the main text which has taken phase transition into consideration, the results are totally different, indicating the necessity of distinguishing PE and FE states in FEM simulation.

Supplementary Figure 8:

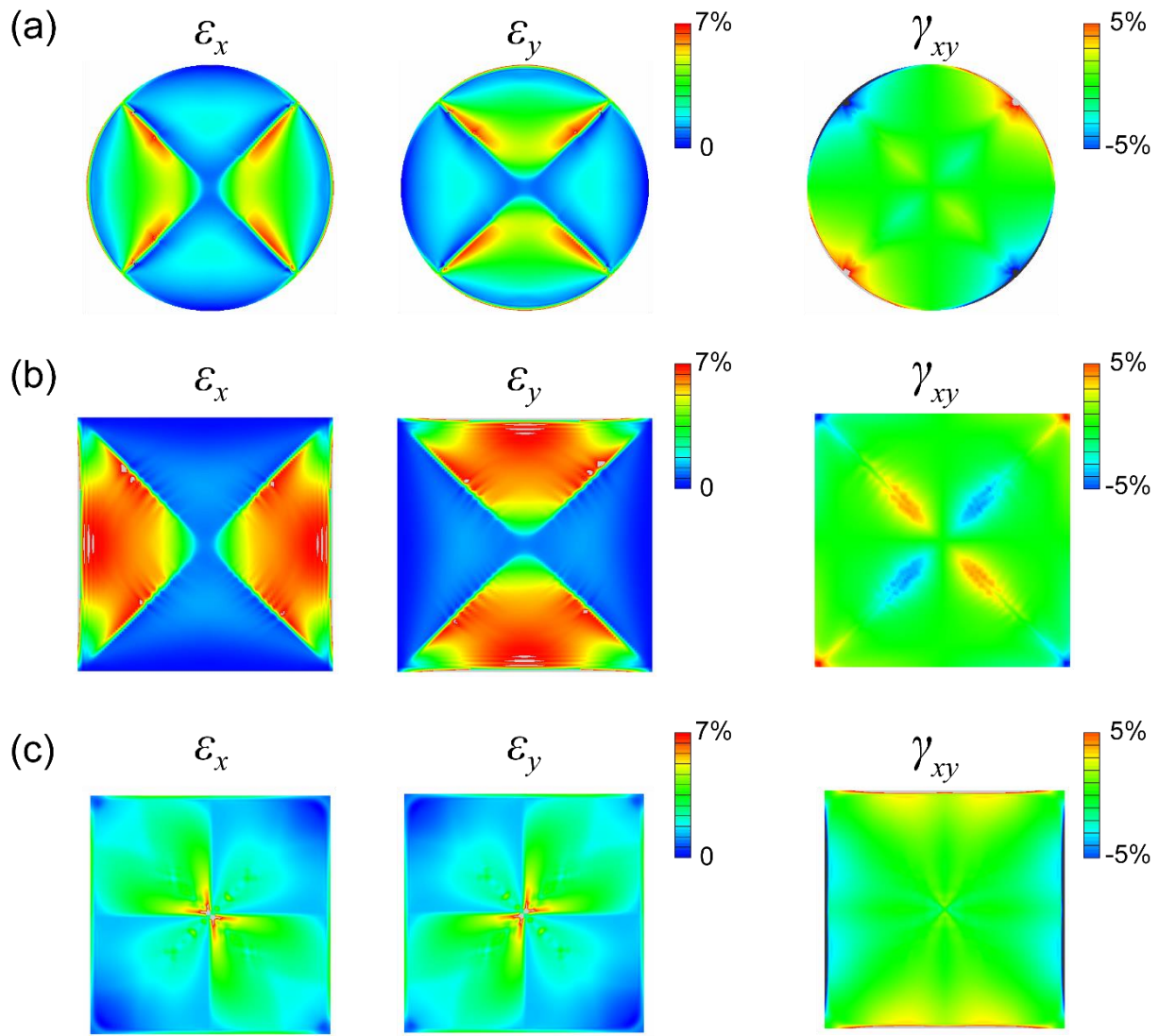

Supplementary Figure 8. The distribution of strain components along the *x*, *y* direction and shear strain in the *x-y* plane of PbTe membrane under three designed loading models. The results here have taken phase transition into consideration, which means the input parameters for FEM simulation here are both the mechanical properties of the PE structure and the FE structure based on the local strain conditions. When the local strain condition is within the PE phase regime, the mechanical parameters of the PE structure are adopted. On the contrary, the mechanical parameters of the FE structure are employed instead when the load strain condition is within the FE phase regime.

Supplementary Figure 9:

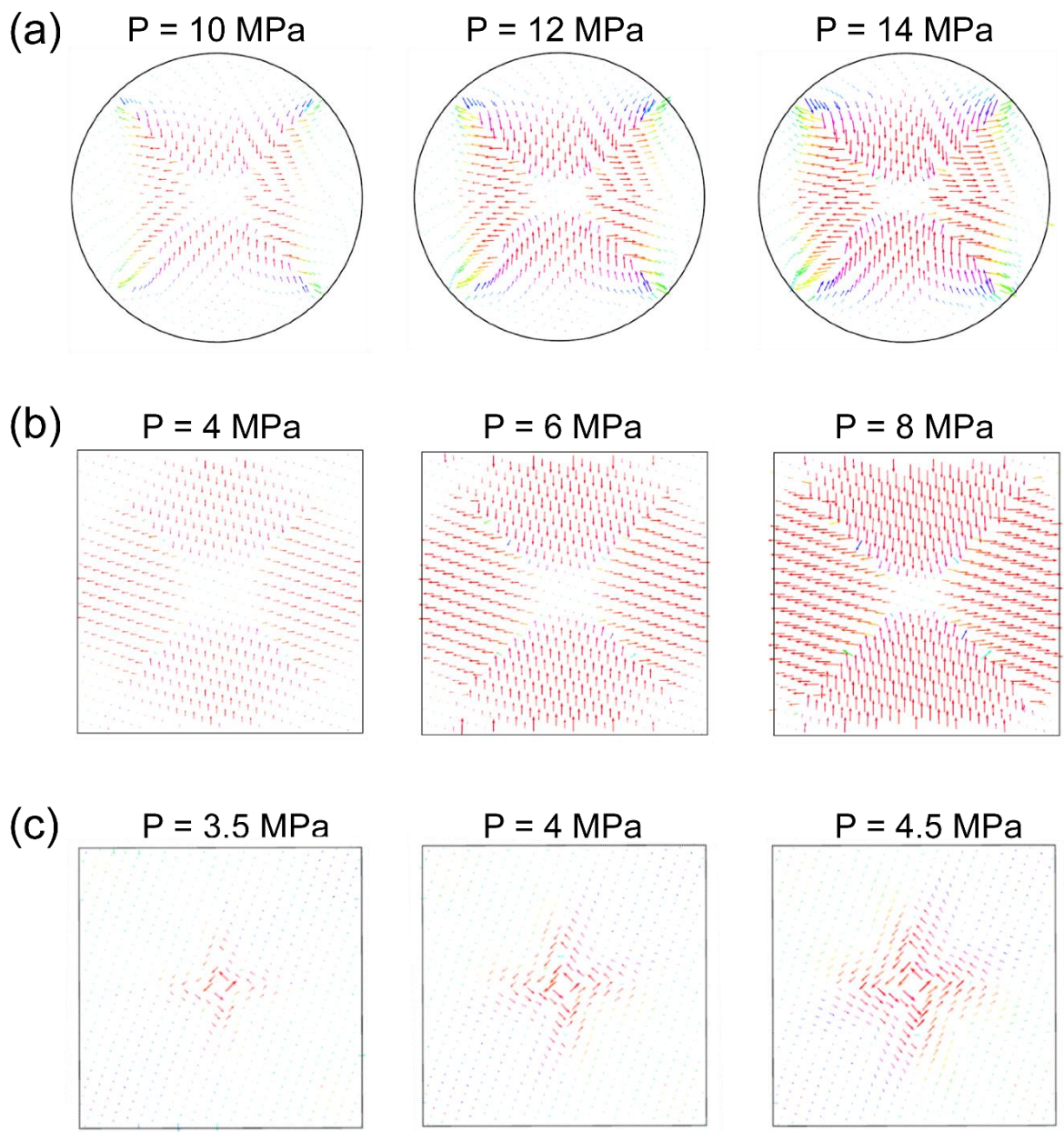

Supplementary Figure 9. The evolution of polar structure under increasing uniform pressure in the three loading models. Different polarization patterns gradually arise with increasing loading. For model I, the final polarization patterns turn into a cubic antivortex polar structure. For model I, the final polarization patterns turn into a round-shape antivortex polar structure. For model III, the final polarization pattern is a flux-closure structure. It means the obtained polarization pattern is strongly dependent on the hole shape of the substrate and the layout of the membrane.

Supplementary Figure 10:

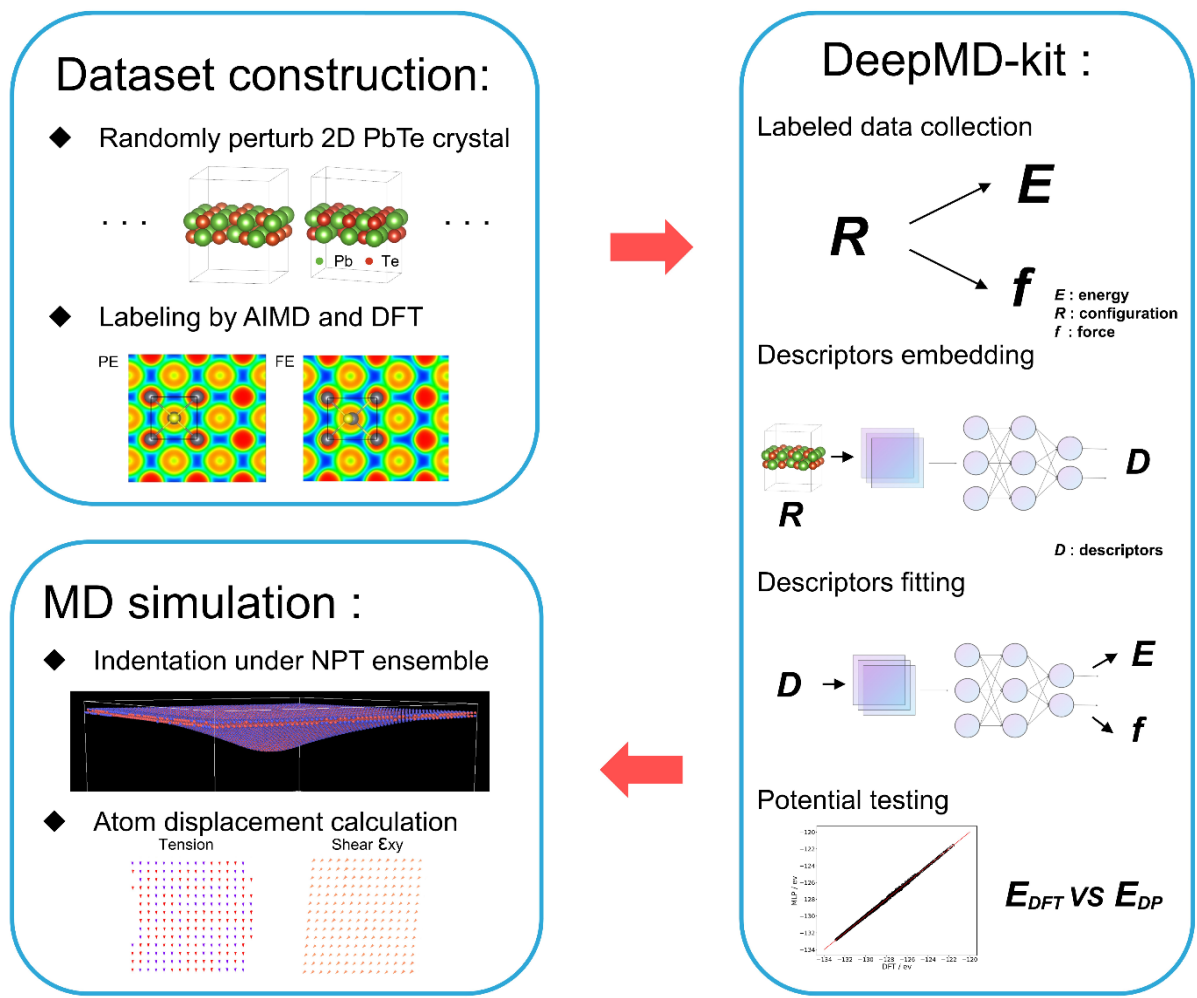

Supplementary Figure 10. The flow chart of building machine learning potential for monolayer PbTe. First, the dataset for machine learning is constructed based on the DFT-based Ab-initio molecular dynamics calculations to obtain a variety of PbTe structures, including the rippled, stretched, and sheared structures. Second, a well-developed DeepMD-kit is employed to train a deep-learning potential for monolayer PbTe. Finally, the obtained potential is used to carry out large-scale MD simulations for mechanical indentation of monolayer PbTe.

Supplementary Figure 11:

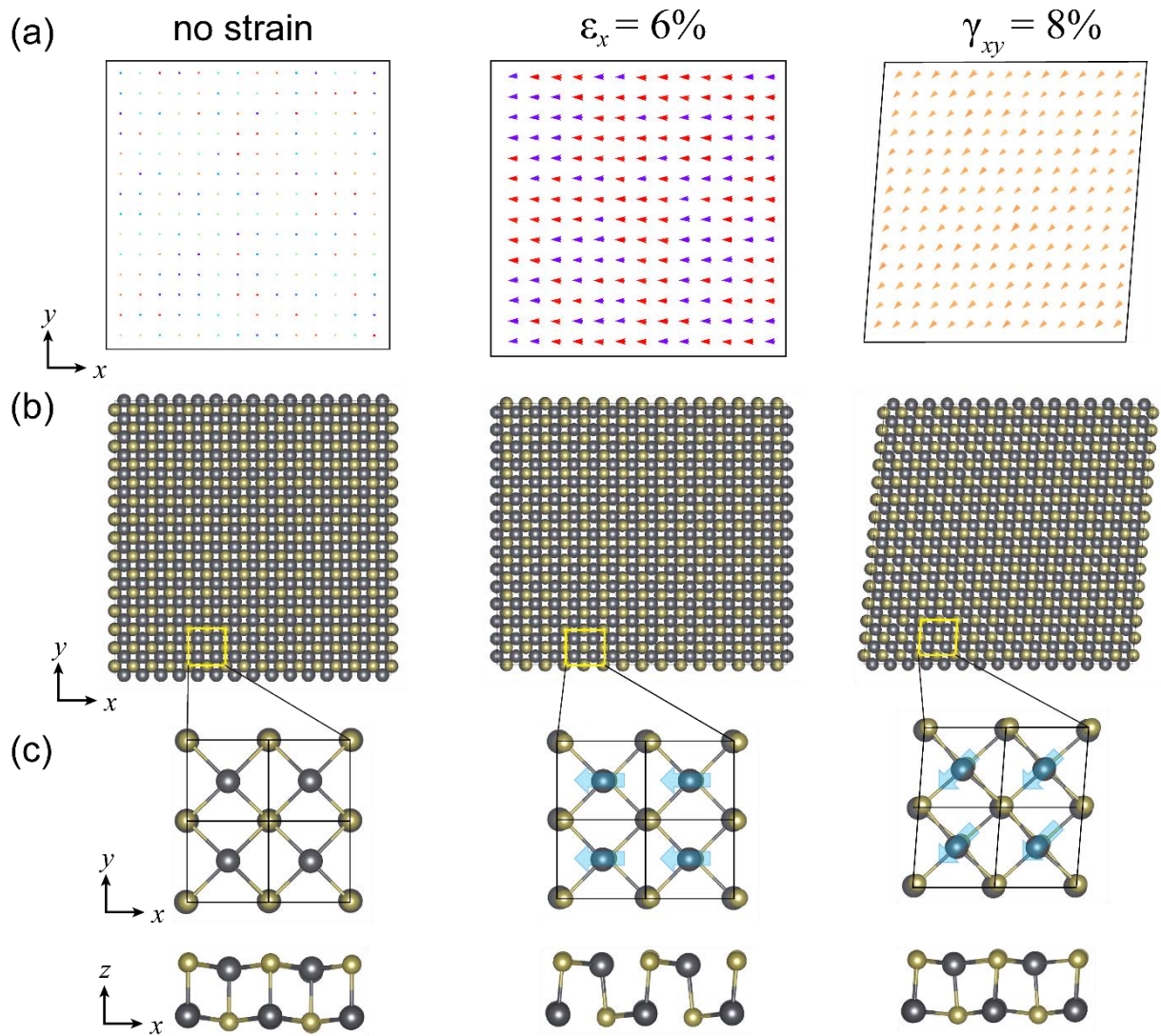

Supplementary Figure 11. The MD simulation results for the polarization and atomic structures of the pristine, stretched, and sheared monolayer PbTe, based on the deep learning potential. It demonstrates that the MD simulations based on DP-potential can reproduce the results of DFT calculations at a larger scale.